----------
X-Sun-Data-Type: default
X-Sun-Data-Description: default
X-Sun-Data-Name: a2634.tex
X-Sun-Content-Lines: 1756

\def \kms{$\rm km\,s^{-1}\,$}

\documentstyle[12pt,aasms]{article}
\lefthead{Scodeggio et al.}
\righthead{Abell 2634 and 2666}

\slugcomment{\it To appear in the Astrophysical Journal}

\begin{document}

\title{The Spatial Distribution, Kinematics, and Dynamics of the Galaxies in
the Region of Abell 2634 and 2666
\footnote{Work based in part on observations obtained at the Arecibo
and Palomar Observatories. Observations at the Palomar Observatory were
made as part of a continuing collaborative agreement between the
California Institute of Technology and Cornell University. The Arecibo
Observatory is part of the National Astronomy and Ionosphere Center,
which is operated by Cornell University under a cooperative agreement
with the National Science Foundation. }}

\author{Marco Scodeggio, Jos\'e M. Solanes, Riccardo Giovanelli,}
\and
\author{Martha P. Haynes}

\affil{National Astronomy and Ionosphere Center and Center for
Radioastrophysics and Space Research,\\
Cornell University, Ithaca, NY 14853\\
e-mail: scodeggi, solanes, riccardo, haynes@astrosun.tn.cornell.edu}
\authoraddr{Center for Radiophysics and Space Research and National Astronomy
and Ionosphere Center, Cornell University, Space Sciences
Building, Ithaca, NY 14853}

\begin{abstract}
A total of 663 galaxies with known redshifts in a $6\deg\!\times 6\deg$
field centered on A2634, including 211 new measurements, are used to
study in detail the structure of the region.  In it we find six main
galaxy concentrations: the nearby clusters A2634 and A2666, two groups
in the vicinity of A2634, and two distant clusters at $\sim 18{,}000$
(A2622) and $\sim 37{,}000$ \kms seen in projection near the core of
A2634.

For A2634, the most richly sampled of those concentrations, we are able
to apply strict cluster membership criteria.  Two samples, ---one
containing 200 galaxies within two degrees from the cluster center and
a second, magnitude-limited, of 118 galaxies within the central half
degree---, are used to examine the structure, kinematics, dynamics and
morphological segregation of the cluster. We show that early-type
galaxies appear to be a relaxed system, while the spiral population
eschews the center of the cluster and exhibits both a multimodal
velocity distribution and a much larger velocity dispersion than the
ellipticals.  We propose that the spiral galaxies of A2634 represent a
dynamically young cluster population.

For the galaxy component of A2634, we find no evidence of significant
substructure in the central regions supportive of a recent merger of
two subclusters, a scenario that has been suggested to explain the
bending of the tails of the cluster central radio source (3C 465). We
also conclude that the adoption of lenient membership criteria that
ignore the dynamical complexity of A2634 are unlikely to be responsible
for the conflictual results reported on the motion of this cluster with
respect to the CMB.

The kinematical and dynamical analysis is extended to A2634's close
companion, A2666, and the two distant background clusters.

\end{abstract}

\keywords{galaxies: clustering --- galaxies: distances and redshifts}

\section{Introduction}

The development of multifiber spectroscopy has made possible the
simultaneous acquisition of tens of galaxy spectra. As a result,
the study of the dynamics of clusters of galaxies is enjoying
the availability of extensive and more complete new redshift data bases, which
reveal the complexity of
detail in cluster structure and help to solve the problems plaguing the
interpretation of the
kinematical data like projection effects, velocity anisotropies, substructure
and secondary infall. In this paper, we present the results of an extensive
observational effort to expand the kinematical data base for A2634, and
a detailed analysis of the structure and dynamics of that cluster.

A2634 is a nearby cluster, classified by Abell (1958) as of richness
class 1.  It was included by Dressler (1980a,b) in his study of the
galaxy populations of 55 nearby clusters and, given its relative
proximity ($z \sim 0.03$), it figures prominently in his cluster sample with
132 galaxies catalogued.  Using Dressler's coarse subdivision into
E, S0, S and Irr types, the fraction of E and S0 galaxies in the central
region of A2634 is approximately $63\%$. This cluster and its close companion
Abell 2666 (around 3$^\circ$ apart and at slightly lower redshift) are
located in a region of complex topology in the background of the
Pisces-Perseus supercluster (Batuski \& Burns 1985; Giovanelli, Haynes,
\& Chincarini 1986a), which further complicates the kinematical analysis.

Matthews, Morgan, \& Schmidt (1964) catalogued the central first-ranked galaxy,
NGC 7720 (UGC 12716), as a cD, thus leading to the classification of A2634
as of Rood-Sastry type cD or Bautz-Morgan type I--II (Sastry \& Rood 1971;
Bautz \& Morgan 1970).
The association of NGC 7720 with the wide-angle tailed (WAT)
radio source 3C 465 (Riley \& Branson 1973) and the absence of a cooling flow
was used by Jones \& Forman
(1984) to cast serious doubts on the possibility that this galaxy was at rest
at the bottom of the cluster potential and to justify the nXD classification of
A2634.
Early radial velocity observations reporting velocity offsets of NGC 7720
with respect to the rest frame of the cluster of more than 300 \kms (Scott,
Robertson,
\& Tarenghi 1977) seemed to support the nonstationary nature of this galaxy
(see, however, Zabludoff, Huchra, \& Geller 1990), although the
significance of these measurements is affected by the small size of the
samples involved. The need to elucidate the process
leading to the formation of the central WAT radio source
has instigated a recent thorough study of the cluster by Pinkney et al. (1993;
hereinafter Pi93). From a sample of 126
redshifts of galaxies within one degree of the cluster center,
Pi93 find a difference between the radial velocities of
the cD galaxy and the whole cluster of $-219\pm 98$ \kms, which they
report as statistically significant. Nevertheless, in their analysis
they also stress that the kinematical properties of the galaxies in the
above sample and the elongation exhibited by the X-ray image of the
central part of the cluster suggest that the northeast region of A2634
may harbor a dispersed subcluster currently undergoing merging with the
primary unit. When they remove this contaminating region, the velocity
offset of NGC 7720 drops to $-85\pm 91$ \kms, a result consistent with
the galaxy being stationary with respect to the cluster primary component. Pi93
then conclude that the bending of the radio tails of 3C 465 is probably
due to large-scale turbulent gas motions fueled by the ongoing merging
process.

The application of the Tully-Fisher (TF) technique (Tully \& Fisher 1977)
to spiral galaxies in nearby clusters, which has led to estimates of $H_0$,
has been used to sample the large-scale peculiar velocity field.
Likewise, the analogous $D_n-\sigma$ relation (Dressler et al. 1987), which is
applied to early-type galaxies, has been extensively applied
to cluster fields. The results of the two techniques have not
always been in agreement. Most notable is the case of the cluster A2634,
where Lucey et al. (1991a) showed that the discrepancy between the
estimates of the peculiar velocity of the cluster as inferred using
the two techniques exceeds 3000 \kms, a value much in excess
of the estimated accuracies of the two methods. The source of this
discrepancy could lie in biases that destroy the universality
of the TF and $D_n-\sigma$ relations, or in problems associated with
the definition of cluster membership. While the issue of universality
for those relations has received significant attention, the problems
arising from misplaced assignment of cluster membership have
not been considered in comparable detail.

A2634 is one of a sample of nearby clusters in each of which we are
obtaining at least 50 redshift-independent distances via both the TF and
the $D_n-\sigma$ methods. Our goals include not only the determination of
a reliable set of cluster peculiar velocities, but also
disentanglement of possible environmental dependences of the TF and
$D_n-\sigma$ relations from the blurring influence of poorly
assessed cluster membership. In this paper, we present 174 new
galaxy redshifts, which we combine with those in the public domain
to carry out a detailed kinematical analysis in the region of A2634 and A2666.
In a forthcoming work, the analysis of the redshift-independent
distances will be presented.

The present paper is organized as follows. In \S~2, we present our new
spectroscopic observations in a
$6\deg\!\times 6\deg$ field centered on A2634, that also includes A2666,
obtained at the Arecibo and Palomar Observatories. In \S~3, we describe
the large-scale characteristics of the region in which A2634 and A2666
are located, outline the main clusters and groups that can be
detected in the $6\deg\!\times 6\deg$ field around A2634
and define strict cluster membership criteria for A2634. In \S~4, we
examine the spatial distribution and kinematics, and their dependence on
morphological type, of the galaxies on a sample that
contains all the cluster members within two degrees of the A2634 center,
and on a more restricted, magnitude-limited, sample within half degree. For
the latter sample, we investigate issues related to the kinematics of
NGC 7720 and the existence of subclustering. In
\S~5 we determine the main kinematical properties of A2666 and the other
clusters and groups around A2634 identified in \S~3. The dynamical analysis of
all
groups and clusters is presented in \S~6. Mass estimates are
given for A2634, A2666 and two background clusters, and the
current dynamical state of the A2634\slash 2666 system is explored. In \S~7,
we summarize the main results.

Throughout the paper we assume $H_0=50$ \kms $\rm Mpc^{-1}$
and $q_0=1/2$. Celestial coordinates are all referred to the 1950 epoch.

\section{New Redshifts}  

The  $6\deg\!\times 6\deg$ field centered on A2634 was visually surveyed
on the Palomar Observatory Sky Survey (POSS) blue prints, identifying
spiral galaxies with major diameter larger than about $0\farcm5$. Coordinates
for
these objects were measured with few arcsec accuracy using
a measuring machine developed by T. Herter.  This search
was carried out to obtain a list of candidates for future TF work. In
addition, the inner square degree of this region was searched
using the FOCAS software package, on digitized images of the Palomar Quick
Survey,
obtained with the kind assistance of D. Golombek of the Space Telescope
Science Institute. This search produced a list of galaxy positions and
rough indicative magnitudes which is estimated to be complete to a limiting
magnitude near 16.5, although numerous fainter galaxies are included.
About half of the galaxies in the first spiral sample were observed with
the Arecibo 305m radio telescope in the 21 cm line. An effort was made
to include all the galaxies in the inner region brighter than 16th
magnitude among those targeted by an optical spectroscopic survey carried
out with the 5m Hale telescope of the Palomar Observatory. Because the
spectra were in great part obtained with a multifiber spectrograph, many
targets were included for reasons of observational expedience rather than
responding to strict flux or size criteria. As a result, the limits of
our spectroscopic survey are blurred, and can roughly be represented by a
completeness
function which, in the inner square degree is 1 at $m_{pg} \sim 15.7$, 0.85
near 16.0, and drops below 0.5 at magnitudes fainter than 16.5.

We report here new redshifts of 174 galaxies in the $6\deg\!\times 6\deg$
field centered on A2634, which also includes A2666. In a companion paper
(Giovanelli et al. 1994), we
present 37 additional redshifts that fall within that region, which were
obtained for a separate but complementary study.

\subsection{Hale 5m Telescope Observations}  

The red camera of the double spectrograph (Oke \& Gunn 1982)
was used on September 25, 1992 to obtain single slit
spectroscopy for 41 galaxies in our catalog. A grating with
$316\rm\, lines\,mm^{-1}$ was used to produce spectra in the range
4900--$7300\rm\,\AA$, with a typical resolution of $9.6\rm\,\AA$. The
detector was a TI $800\times 800$ CCD chip.  The data were reduced
following standard procedures with available IRAF\footnote{IRAF
(Image Reduction and Analysis Facility) is distributed by the National
Optical Astronomy Observatories, which are operated by the Association
of Universities for Research in Astronomy, Inc., under contract with the
National Science Foundation.} tasks.  Each frame was overscan- and
bias-subtracted,
then flat-fielded using appropriate dome flats. Wavelength
calibration was performed using He-Ne-Ar lamp comparison spectra before
and after each source observation.
Heliocentric velocities were computed in two different ways: using the
IRAF task {\it fxcor\/}, which is based on the cross-correlation algorithm
described by Tonry \& Davis (1979), for pure absorption spectra, and
measuring the central wavelength of the gaussian fit to the emission
lines of spiral galaxies (typically between 3 and 7 lines were measured
for each galaxy). Three different K giant stars were used as templates
for the cross-correlation.

The Norris multifiber spectrograph (Hamilton et al. 1993) was used on
October 20, 1993 to observe a fainter sample of galaxies in the central
region of A2634 and A2666.  The Norris spectrograph has a field of view
of $20\arcmin$, within which 176 independent fibers can be placed for
photon acquisition. At the front end, the fibers subtend $1\farcs6$ and
can be placed as close as $16\arcsec$ apart. The fiber outputs are
transferred through a grating element to a CCD detector. At the time of
these observations,
the Tektronik CCD had a reduced format of $1024\times 1024$, restricting the
number of usable fibers to half the total and the field of view to a
$10\arcmin\times 20\arcmin$ region. Coordinates for the
target galaxies were measured on the digitized Palomar ``quick V''
survey plates at STScI\footnote{Astrometry was obtained using the Guide
Stars Selection System Astrometric Support Program developed at the
Space Telescope Science Institute, which is operated for NASA by the
Association of Universities for Research in Astronomy, Inc.}. A
$300\rm\,lines\,mm^{-1}$ grating was used to produce spectra in the
range 4500--$7000\rm\,\AA$, with a  resolution of $7.5\rm\,\AA$. Each frame
was first overscan- and bias-subtracted, then one-dimensional spectra were
extracted using appropriately sized apertures. These spectra were
wavelength-calibrated using  He-Ne-Ar comparison lamp spectra taken with the
fibers deployed as for the galaxy acquisition, and extracted using the
same apertures used for the object spectra. The sky subtraction was
performed subtracting from each one of the object spectra the median of
the output of ten fibers positioned on blank sky positions, properly
scaled in order to compensate for light transmission variations from
fiber to fiber. After the final one-dimensional spectra were obtained,
velocities were derived with precisely the same procedure described
above for the single slit spectra. Two different K giant stars were
used as templates for the cross-correlation.

The optical radial velocities and their associated errors are listed in
Tables 1 (longslit data) and 2 (multifiber data).  In column (1) the
CGCG (Zwicky et al.  1963--68) identification (field number and ordinal
number of galaxy in the field), or NGC number is given, when available.
Otherwise,  the galaxy name follows an internal catalog number
designation.  In columns (2) and (3), the right ascension and
declination are listed; columns (4) and (5) give the heliocentric
radial velocity and its associated error, as estimated combining the
uncertainty in the calibration and the one derived from the measured
signal to noise ratio of the cross-correlation function (``a'' in
column [6]), or estimated from the dispersion in the single emission
line velocity determinations, combined with the
calibration uncertainty (``e'' in column [6]), both in  \kms. Column (7)
lists the velocities found in the literature for some of these
galaxies, with the references coded in column (8).

\subsection{Arecibo Observations}  

New $21\rm\,cm$ observations of 89 galaxies in the A2634\slash 2666
region were carried out between 1990 and January 1994. In all cases the
observational setup was as described in Giovanelli \& Haynes (1989).
Since this sample includes galaxies significantly fainter than those
customarily observed in $21\rm\,cm$ emission, integration times
averaged 0.7 hours per object on source. Typical rms noise per averaged
spectrum ranged between 0.3 and $0.9\rm\,mJy$. The later set of
observations (July 1993 to January 1994, approximately $25\%$ of all
runs) benefited from the completion of Phase I of the Arecibo telescope
ongoing upgrading project (Hagfors et al. 1989), which includes the
construction of a ground screen, 15 meters high and covering an area of
$14{,}700\rm\,m^2$ around the main dish. The principal benefit
of this device is a drastic reduction in the amount of ground radiation
pickup, allowing a significant
reduction of system temperature and an increase in the telescope
sensitivity for observations with zenith angles higher than $10\deg$.
All observations were taken with a spectral resolution of approximately
8 \kms, later reduced by smoothing by an amount dependent on the
signal-to-noise ratio. Table 3 lists the galaxies for which new $21\rm\,cm$
spectra were obtained. In columns (1) and (2), the CGCG (Zwicky et al.
1963--68) identification (field number and ordinal number of galaxy in
the field) or the NGC/IC number, and the UGC (Nilson 1973) number are
given, when available. Otherwise, in column (1) the galaxy name follows an
internal catalog number designation. The right
ascension and declination are in columns (3) and (4); columns (5) and (6)
give the heliocentric $21\rm\,cm$ radial
velocity and its associated error, estimated from the signal-to-noise
ratio and structural parameters of the line profile, both in  \kms.
For those galaxies in common with Pi93 we list in column (7) the velocities
given by
these authors.
About one third of the entries in Table 3 had been previously observed
by us at Arecibo, but data were of inferior quality. Listings in Table 3
supersede previous
reports, which are coded in column (8).

\subsection{Comparison with Previous Work} 

Optical observations for 20 galaxies and $21\rm\,cm$ line observations
for 15 overlap with the list of observations presented by Pi93. The
comparison between these two sets of redshift determinations reveals a
general good agreement and few large discrepancies, which usually
involve the measurements affected by the largest uncertainties. As
already pointed out by Pi93, in the case of low signal to noise ratio
spectra (like their galaxies in the C2 category), the cross-correlation
function has typically more than one peak, because of possible chance
alignments of real and noise features in the object and template
spectra. The choice of the wrong peak might then result in redshift
determinations with errors arbitrarily large. We suscribe to the
cautionary words of Pi93 about the redshift measurements inferred from
the noisiest spectra, and notice that, likewise,
velocities in Tables 1 and 2 with associated errors on the order of
200 \kms or larger should be considered as doubtful entries.  In the
following analysis the velocities listed in column (4) of Tables 1 and
2, and in column (5) of Table 3, are used for all the galaxies with
multiple observations.

\section{The Environment of A2634 and Cluster Membership}  

\subsection{The Large-Scale Environment}  

The cluster A2634 appears projected on the main ridge of the
Pisces-Perseus supercluster (hereinafter referred to as PPS; see Fig.~1 of
Wegner,
Haynes, \& Giovanelli 1993), although it and its close companion A2666 are
actually located on the higher redshift branch of the two into which the
PPS splits near $\rm R.A. = 00^h 45^m$. Figures 1 and 2
illustrate respectively the projected spatial distribution and the
distribution in redshift space of the galaxies with measured radial
velocities in our PPS catalog, pertaining to a region bounded by $\rm
22^h\le R.A.\le 1^h$, $\rm 20\deg \le Dec.\le 35\deg$ and $0\le
cz_{hel}\le 16{,}000$ \kms, which well describes the large-scale
environment around A2634. The radial
velocities included in Fig.~2 come from Giovanelli \& Haynes
(1985, 1989, 1993), Giovanelli et al. (1986b), Wegner et al. (1993),
Pi93, this paper, and Giovanelli et al. (1994).
A2634 is the most conspicous galaxy concentration dominating the center of
Figs.~1
and 2; it is located in a region of high galactic density at a distance of
$\sim 9000$ \kms. Approximately $3\deg$ to
the east is the lesser concentration represented by A2666, with the PPS
ridge extending to the northeast. Several galaxy groups appear also
prominent in the same figure.
The substantial galaxy concentration associated with A2634 is partly due to the
fact that in the cluster region radial velocities are measured to a fainter
flux limit than in the rest of the field. It is clear from Figs.~1 and 2
that the topology of the region surrounding
A2634 is quite intricate, perhaps more so than any in the PPS region.

Next, we concentrate on the more immediate neighborhood of A2634,
namely the $6\deg\!\times 6\deg$ field around its center where our new
redshifts have been measured. Figure 3 displays the positions of
663 galaxies with known redshift in that region. Of those, 237 are
redshifts obtained from our previous surveys, 215 are listed in
Pi93 and 211 are the new determinations presented in this paper and in
Giovanelli et al. (1994). In Fig.~3, filled circles are used for objects with
$m_{pg}\le 15.7$ for which the redshift survey is virtually complete
in the whole field, while open circles are used
for galaxies fainter than that limit. Note that the latter are clearly more
concentrated towards the cluster inner regions due to the
observational bias towards the core of A2634 mentioned before.  In
Figure 4 a radial velocity histogram between 0 and $45{,}000$ \kms
of the galaxies in Fig.~3 is shown: the significant peak near 5500 \kms is
associated
with the foreground branch of the PPS; the peak near 8000 \kms includes
principally galaxies in the redshift domain of A2666; the highest
peak near 9000 \kms is associated with A2634; and a well defined group around
$11{,}700$ \kms appears detached from the A2634 regime.
The feature around $18{,}000$ \kms includes the
rich Abell cluster A2622 (centered $0\fdg9$ to the NW of A2634)
and several more widely spread galaxies.
The velocity peak near $37{,}000$ \kms is produced by a noticeable
concentration of galaxies almost directly behind the A2634 center.
Pi93 tentatively associate the X-ray clump seen in the
{\it Einstein\/} IPC map of A2634 to the northwest of  A2634 itself
(see their Fig.~4) with this background
distant cluster. In \S~5.2, we analyze this possibility in more detail.

\subsection{Cluster Membership}  

As a first quantitative step in our analysis, we have assigned
cluster/group membership to the galaxies in our $6\deg\!\times 6\deg$
sample.  Usually, cluster membership is assigned using only a fixed range of
velocities.
The velocity distribution of all galaxies projected within
a chosen distance from the cluster center is obtained and field galaxy
contamination is eliminated, typically via a 3$\sigma$-clipping
technique (Yahil \& Vidal 1977) or using procedures that take
into account the gaps of the observed velocity distribution. At this
point, all galaxies within two fixed velocity limits are assigned to the
cluster. These methods do not consider the fact that, with the exception
of the virialized inner cluster regions where a roughly constant velocity
dispersion is expected, the line-of-sight velocity dispersion of true cluster
members decreases with increasing projected separation from the cluster center,
whether the cluster is bound and isolated or whether a significant
amount of secondary infall is present. This consideration will be taken into
account in the subsequent membership assignment for the best sampled cluster
A2634.

Because of the proximity between A2634 and A2666, we
first proceed to identify those galaxies in the neigborhood of A2634
that are more likely to be associated with A2666. To accomplish that,
we compare the projected number density profiles of the
two clusters at the position of any given galaxy, and assign it to A2666 if
this cluster's density profile has a higher value than that of A2634 at that
location. The projected number density profile of A2634 has been
determined applying the so-called ``direct method'' of Salvador-Sol\'e,
Sanrom\'a, \& Gonz\'alez-Casado (1993a) to the Dressler's (1980b)
magnitude-limited sample of A2634 galaxies (see \S~4.3 below for
details), which covers the inner region of the cluster up to a
radial distance of $\sim 0\fdg5$ from its center (within this region a
negligible contamination by galaxies belonging to A2666 is expected as
the two clusters are separated by $3\deg$).
A good fit to the numerical solution given by this method is
obtained using a modified Hubble-law of the form
\begin{equation}\label{den_2d}  
N(X)=N(0)[1+X^2]^{-\gamma+1/2}\;,
\end{equation}
where $N(0)$ is the central projected number density of galaxies,
$\gamma$ a shape index, and $X=s/r_\ast$ the projected distance $s$
to the center of the cluster in core radius units $r_\ast$. The best
fitting values of the parameters $N(0)$, $\gamma$ and $r_\ast$ for A2634
are:  $69\rm\,galaxies\,Mpc^{-2}$, 1.5, and $0.48\rm\,Mpc$,
respectively. For A2666, the scarcity of observations makes the
above procedure more uncertain; we then assume, for simplicity, that
the values of the shape index and core radius of both clusters are the
same and that they have the same mass-to-light ratio.
Next, for each
galaxy in the sample with heliocentric velocity in the range
6500--9500 \kms (which we adopt as the velocity boundaries of A2666; see
\S~5.1), we compute the projected distances to A2624, $X_{34}$, and
A2666, $X_{66}$, and assign A2666 membership if $N (X_{66}) > N
(X_{34})$ or, equivalently, if
\begin{equation}  
X_{66} < \biggl[{M_{66}\over{M_{34}}}(1+X_{34}^2
)-1\biggr]^{1/2}\;,
\end{equation}
where $M_{66}/M_{34}$ is the ratio of the masses of the two clusters,
which we take equal to $1/7$ (see \S~6.1). For the center of A2634 we adopt
the peak of the X-ray emission $\rm R.A. = 23^h 35^m 54\fs 9$, $\rm
Dec. = 26\deg 44\arcmin 19\arcsec$ (C. Jones 1993, private
communication), while for A2666 we use the peak density of the galaxy
distribution $\rm R.A. = 23^h 48^m 24\fs 0$, $\rm Dec. = 26\deg
48\arcmin 24\arcsec$. To convert angular separations to physical units,
we calculate the cosmological distances of the clusters from the
redshifts in the Cosmic Microwave Background (CMB) reference frame (see
\S~4.2 and 5.1): $z_{CMB} (\rm A2634) = 0.0297$ and $z_{CMB} (\rm A2666) =
0.0256$. It is worthwhile to mention that the exact values adopted for
the ratio of cluster masses (or, equivalently, the ratio of central
projected number densities), $\gamma$ and $r_\ast$, have a negligible
practical influence on the results to be discussed later on, so any
further refinement of this procedure is unnecessary.

In Figure 5 we plot the heliocentric velocity as a function of the
angular separation $\theta$ for all galaxies within $6\deg$ of the
A2634's center.  Unfilled symbols identify probable A2666 members using
the procedure described above. Since contamination from A2666 starts to
be present at $\theta\gtrsim 2\deg$, we choose to restrict our
following analysis to galaxies within this distance from the center of
A2634. The solid lines plotted in Fig.~5 are the caustics
associated with A2634 derived using the spherical infall model described
in Reg\"os \& Geller (1989). In
this simple model, the amplitude of the caustics depends only on $\Delta(r)$,
the mass overdensity profile of the cluster, and the cosmological
density parameter $\Omega_{0}$. To determine the former, we have
followed the procedure outlined by Reg\"os \& Geller (1989), which
relies on the assumption that the galaxies trace the mass distribution,
i.e., that $\Delta(r)$ is identical to the {\it number\/} density
enhancement of galaxies in the cluster:
\begin{equation}\label{den_enh}  
\Delta(r)={<\rho>\over \rho_u}-1={<n>\over n_u}-1\;.
\end{equation}
In equation~(\ref{den_enh}), $<\rho>$ and $<n>$ are, respectively, the
average mass and number density of cluster galaxies (corrected for
field contamination) inside a radius $r$, whereas $\rho_u$
and $n_u$  are the mean mass density and the mean number density
of galaxies in the universe. The spatial distribution of galaxies $n(r)$ has
been estimated by inverting the analytical fit to the observed projected number
density profile $N(r)$ given by eq.~(\ref{den_2d}). The mean number density
of field galaxies $n_u$ has been calculated by integrating the Schechter (1976)
luminosity function obtained from the CfA survey (de Lapparent, Geller, \&
 Huchra 1989) to the limiting magnitude of the Dressler sample, $m_v=16.0$.
This has been converted to the $B(0)$ scale of the CfA survey
assuming $m_{B(0)}-m_v=0.8$. We obtain
$n_u=2.28\times 10^{-3}\rm\,galaxies\,Mpc^{-3}$.

The caustics of Fig.~5 correspond to $\Omega_{0}=0.2,
0.5$, and 1, the amplitude increasing with the value of $\Omega_{0}$.
Clearly, no sharp boundary is observed in the velocity distribution of
the galaxies in this sample that would mark the transition between the
triple-valued and single-valued regions, which suggests that some of the
simplifying assumptions implicit in the
Reg\"os \& Geller model are probably too stringent (see,
for instance, van Haarlem et al. 1993). But the caustics derived from the
spherical infall model can
still be used effectively to accomplish our purpose of defining strict cluster
membership if they are supplemented
by further inspection of both the spatial and velocity distribution
of candidate A2634 members. Figure 6a shows an enlargement of Fig.~5,
focusing on galaxies in the inner $2\deg$ around the center of
A2634, while Figure 6b displays their corresponding sky
distribution. Two groups of galaxies between the
$\Omega_{0}=0.5$ and $\Omega_{0}=1.0$ caustics can be separated
from strict cluster members: one group probably
located in the foreground of the cluster (hereinafter
A2634--F) between 7000 and 8000 \kms, and a second in the background
(hereinafter A2634--B) between $11{,}000$ and $12{,}000$ \kms already
identified as a detached group by Pi93.
Both groups have small velocity dispersions, are separated by large
gaps in radial velocity from other galaxies at a comparable distance from
the cluster center and appear to be concentrated in small areas in the sky,
to the east of the A2634 core (approximately $1\fdg5$ to the southwest
of the center of A2634 there is a possible third group of foreground
objects with velocities around 7000 \kms, but the evidence for its physical
reality is poor). We assign separate dynamical identity to these two groups;
additional justification for this choice is presented in \S~6.2. Accordingly,
we chose to consider {\it bona fide\/} A2634 members only those galaxies
inside the $\Omega_{0}=0.5$ caustic. We caution again that this
choice represents more an operational criterion for defining cluster membership
than a statement on the value of $\Omega_0$.

\subsection{Implications of Cluster Membership on Peculiar Motion Measurements}

Lenient assignment of cluster membership and systematically different
kinematic behavior among galaxies of different morphology may be
factors in producing conflicting estimates of the peculiar velocity of
a cluster with respect to a reference frame comoving with the Hubble
flow. A2634 constitutes a glaring example of conflicting results:  the
peculiar velocity of the cluster as inferred by Lucey et al.  (1991a)
by applying the $D_n-\sigma$ technique to a sample of 18 early-type
galaxies is $-3400\pm 600$ \kms, while that inferred from a sample of 11
spirals by Aaronson et al. (1986) applying the TF technique is
essentially zero.  In their analysis, Lucey et al. noticed that the
Aaronson et al.  sample could be divided into two velocity subsets, one
associated with A2634 with 6 members and a CMB systemic velocity of
$8692\pm 92$ \kms, and the other associated with A2666 with 5 galaxies
and $V_{CMB}=7195\pm 349$ \kms, but the TF distances they inferred for
these two subsets still implied negligible peculiar motions. We note,
however, that according to the membership criteria discussed in \S~3.2
only 5 galaxies of the Aaronson et al. late-type sample are proper
members of A2634, while 2 are peripheral members (they are within the
$\Omega_0 = 0.5$ caustic, but at a distance greater than $2\deg$ from
the A2634 center), 2 are members of the A2634--F group and 2 are A2666
members. Clearly, the situation on the TF side is far from being
satisfactory. On the other hand, the origin of the discrepant
$D_n-\sigma$ distance cannot be fully attributed to the presence of
separate dynamical units in the region, or to ill-defined membership
criteria, because all 18 early-type galaxies (9 E's and 9 S0's) used by
Lucey et al. (1991a) for their distance determination are all A2634
members. A full discussion about other possible origins of that
discrepancy is beyond the purpose of this paper. However, we point out
that a surface brightness bias similar to the one observed in the
elliptical population of the Coma Cluster by Lucey, Bower \& Ellis
(1991b), and discussed and dismissed for A2634 by Lucey et al. (1991a),
is not only expected in a cluster at a distance comparable to A2634,
but should be large enough to originate a spurious peculiar motion of
precisely the magnitude derived by these authors for A2634.  Better
understanding of this complicated scenario should be provided by TF and
$D_n-\sigma$ distance determinations based on the larger samples of
strict cluster members being currently analyzed.

\section{Kinematics and Spatial Analysis of A2634}  

Following the recommendations of Beers, Flynn, \& Gebhardt (1990), we
characterize the velocity distribution of the A2634 galaxies by means
of the biweight estimators of central location, $V$ (i.e.,
systemic velocity, sometimes in the literature referred to as $C_{\rm BI}$),
and scale, $\sigma$ (i.e., velocity dispersion, sometimes in the literature
referred to as $S_{\rm BI}$), because of their robustness (i.e.,
insensitivity to the probabilistic model adopted for the observed data)
and resistance (i.e., insensitivity to the presence of outliers). The
errors associated with these estimates, $\epsilon_V$ and $\epsilon_\sigma$
respectively, correspond to $68\%$ bootstrap confidence
intervals based on $10{,}000$ resamplings. The program ROSTAT, kindly
provided by T. Beers, is used for all these calculations.
Among the wide variety of statistical tests implemented in this program to
assess the gaussianity of the empirical distribution, we quote here
the results for the $W$-, $B_{1}$-, $B_{2}$-, and $A^2$-tests
(definitions of these tests can be found, for instance, in Yahil \& Vidal 1977
and D'Agostino 1986). The scaled tail index TI
(Rosenberger \& Gasko 1983; see also Bird \& Beers 1993) is applied to
determine the amount of elongation of the empirical distribution
relative to the gaussian (which is neutrally elongated by definition:
TI=1.0). Distributions which have more strongly populated tails than the
gaussian have TI$>$1.0, while those with underpopulated tails have TI$<$1.0.
Finally, we also investigate the existence in the velocity distribution of
weighted gaps of size 2.75 or larger, corresponding to a probability of
ocurrence in a normal distribution of less than $1\%$ (Beers et al.
1990). We refer the reader to the listed references for a detailed
explanation of these statistical techniques.

If the statistical sample is large enough, the analysis of the possible
correlation between galaxy morphology and kinematics provides a tool
that is helpful in assessing the dynamical and evolutionary state of a cluster
of galaxies. Several studies (e.g., Kent \& Gunn 1982; Tully \& Shaya 1984;
Sodr\'e, Capelato, \& Steiner 1989) show evidence for a higher velocity
dispersion of the spiral population than that of the early-types in
clusters. This result is consistent with the idea that the spiral
galaxies are currently infalling and have not suffered appreciable
dynamical mixing with the virialized core.  Should this picture turn
out to be correct, it would imply that either the spiral fraction in
clusters is increasing with time, and/or that environmental effects are
an important factor in determining galaxy evolution and the relation
between morphology and density, transforming spiral galaxies into
earlier morphological types.  For this reason, we split each one of the
samples selected for A2634 into two subsets of early- and late-type
galaxies and comparatively investigate their kinematics and spatial
distribution.

The results of the kinematical analysis presented in this and in the
following sections for all the clusters within the $6\deg\!\times
6\deg$ field are summarized in Table~4.  The name of the cluster and of
the subsample considered are listed in columns (1) and (2), and the
number of galaxies in the sample is in column (3).  Columns (4)--(7)
list the heliocentric systemic velocity $V_{hel}$, the dispersion
$\sigma$ and their associated $68\%$ boostrap errors. Column (8) gives
the value of the cluster Abell radius $r_A$ in degrees calculated using
cosmological distances. Columns (9) and (10) give the masses of the
various clusters, as obtained in \S~6.1.

All the relevant statistical results for the same samples and their
associated probabilities, are summarized in Table 5. In column (1) we
list the name of the sample and in column (2) the number of galaxies in
it. Columns (3)--(10) list the values of the test statistic and
associated significance levels for the $W$-, $B_{1}$-, $B_{2}$- and
$A^2$-tests, respectively. The significance levels refer to the
probability that the empirical value of a given statistic could
have arisen by chance from a parent gaussian distribution, i.e., the
smaller the quoted probability the more significant is the departure
from the null hypothesis. We choose to discuss only those cases where
the gaussian hypothesis can be rejected at better (i.e., lower) than
the $10\%$ significance level.  Column (11) reports the value of the
scaled tail index TI. The number of weighted gaps of size equal to or
larger than 2.75 and the cumulative probability of finding these many
highly significant weighted gaps somewhere in the distribution are
listed in columns (12) and (13), respectively.

\subsection{The Two Degree (TD) Sample}  

After the contamination due to nearby groups has been removed via
the procedure described in \S~3.2, our sample of bona fide
A2634 members within a radius of 2 degrees (hereinafter referred to as the ``TD
sample'') is composed of 200 galaxies.
The stripe density plot of heliocentric radial velocities and the
corresponding histogram for this sample are shown in Figure~7.
For this sample $V_{hel}=9249^{+55}_{-56}\,$ \kms and
$\sigma =716^{+48}_{-35}\,$ \kms after applying relativistic
and measurement error corrections (Danese, De Zotti, \& di Tullio
1980). In the direction of A2634, the motion of the Sun with respect to the
Local
Group (LG) of galaxies gives a correction
$\Delta V_{LG}=V_{LG}-V_{hel}=244$ \kms, while the corresponding
correction to the CMB rest frame (Smoot et al. 1991) is
$\Delta V_{CMB}=V_{CMB}-V_{hel}= -364$ \kms.  At the cosmological
distance of A2634, one $r_A$ corresponds to 1.02 degrees.

The visual inspection of Fig.~7 yields suggestive but inconclusive
indication of deviation of the velocity distribution from gaussianity.
The dictum of statistical tests, unsurprisingly, is equally ambiguous.
The results of the $B_{1}$-, $B_{2}$- and $A^2$-tests
do not indicate significant departures from normality. However,
the sensitive $W$-test gives only a $7\%$ probability that the observed
distribution
could have arisen by chance from a parent gaussian population.
The tail index TI indicates slight contamination of the tails of the
distribution with respect to the gaussian. The velocity distribution also
contains
one highly significant weighted gap with a {\it per gap\/} probability of 0.006
(the cumulative probability is not statistically
significant), indicated by the arrow in Fig.~7. The size of the corresponding
gap is 83 \kms. The presence of this gap in the observed distribution
suggests, ---but only weakly---, that
the underlying parent distribution may be kinematically complex.

For the TD sample, its subdivision into two
subsets of early- (E and S0) and late-type (S and Irr) galaxies
reveals that the kinematical characteristics of the two populations are
different. Figure~8a shows the stripe density plot of radial velocities
and histogram for the early-type subsample, which has
$V_{hel}=9276^{+67}_{-73}\,$ \kms and $\sigma=658^{+62}_{-42}\,$ \kms.
Despite the low value of the tail index (TI=0.91) and the existence
of one highly significant gap of size 94 \kms (identified with the arrow in
Fig.~8a) near 9000 \kms that weakly suggest possible bimodality in
the parent distribution, none of the gaussianity tests reveals inconsistency
with a normal parent population. On the contrary, the velocity distribution of
the late-type subsample represented in Figure~9a is clearly non-gaussian (the
most
sensitive tests of normality, $W$ and $A^2$, reject the gaussian
hypothesis at better than the $3\%$ level of significance) and bimodal.  Two
highly
significant and very close gaps of 334 and 192 \kms, with {\it per gap\/}
probabilities of only 0.0005 and 0.002, respectively, can be seen
at both sides of two galaxies with radial velocities of 8500 \kms,
splitting the distribution of velocities into two modes which peak around
7700 and 9300 \kms and are responsible of the high value of the tail
index (TI=1.23). The 7700 \kms mode contains four
times fewer galaxies than the 9300 \kms mode, for which location and
scale are approximately coincident with those of the velocity
distribution of the early-type galaxies. This result implies that the
low-velocity tail seen in the velocity histogram of the TD sample (Fig.~7) is
largely dominated by late-type galaxies (notice that the 10 likely nonmembers
discussed in the Pi93 study have already been removed from the present sample,
as they all belong to the A2634--B and A2634--F groups).
Although a two-sample Kolmogorov-Smirnov (KS)
test gives a probability of $25\%$ for the null hypothesis that early-
and late-type subsamples are drawn from the same parent kinematical
population, the systemic velocity of the late-type galaxies
$V_{hel}=9092^{+99}_{-126}\,$ \kms
is marginally inconsistent, within the adopted uncertainties, with
that of the early-type subsample. In the next section we shall see that the
distributions of early- and late-type galaxies in the central regions of the
cluster
are even more markedly different. The results of the statistical analysis of
these
two subsamples are summarized in Table 5.

The spatial distributions of the galaxies in the two subsamples are
shown in Figures 8b and 9b, with coordinates measured with respect to the
center adopted for A2634 (see \S~3.2). Crosses identify galaxies
with radial velocities less than 8700 \kms, while open circles identify those
galaxies
with velocities equal or above this limit. Although there is no noticeable
spatial
segregation among the galaxies belonging to each one of these two velocity
subgroups,
the comparison of the spatial distributions of the two subsamples reveals two
remarkable aspects. First, and in agreement with the
morphology/density (Dressler 1980a) and morphology/clustercentric
distance relations (Whitmore \& Gilmore 1991), the early-type subset is
strongly concentrated towards the center of the cluster (median radial
distance of 0.24 degrees), while the late-type galaxies are less
strongly so (median radial distance of 0.45 degrees). In addition, the spatial
distribution of the early-type population reveals an apparent scarcity of these
galaxies
at distances larger than $30\arcmin$ from the cluster center along the
north-south
direction, while that of the late-type population does not show any noticeable
asymmetry. One possible interpretation of this apparent elongation is that the
cluster has suffered a recent merger. Indeed, this possibility has been already
suggested by Pi93 who brought attention to the elongation of the ICM, in a
direction consistent with that of the galaxy component (see also Eilek et al.
1984 and our Fig.~17).
These authors claim that the rough alignment of the position angle of
the X-ray image with the direction of the axis of symmetry of the WAT
radio source and the lack of a cooling flow
may be explained by the recent ocurrence of a cluster-subcluster
merger along a line contained in the plane of the sky. In this scenario, the
quasi-stationary WAT is shaped and powered by the
the ICM of the merging subunit, which provides the high relative velocities
($\gtrsim 1000$ \kms) required by the ram pressure model for the bending to be
possible.
We point out, nonetheless, that for clusters immersed in a very large
supercluster structure like A2634, the tidal interaction caused by the
supercluster may also account for the observed elongation of the galaxy
component and the ICM, and their alignment with the first-ranked galaxy
(Salvador-Sol\'e \& Solanes 1993).

In the next three sections, we will study in detail the dynamical
status of the galaxy component of A2634 and apply a series of
statistical tests for the detection of significant spatial and
kinematical substructure that can add further support to the merger
hypothesis. Since an adequate study of substructure in clusters
requires magnitude-limited samples with a substantial number of
galaxies and free from field contamination, we will concentrate our
subsequent analysis on the Dressler's (1980b) magnitude-limited sample
of the central region of A2634.

\subsection{The Half Degree (HD) Sample}  

Within the inner $0\fdg5$ from the A2634 center
(approximately 0.5 $r_A$), Dressler (1980b) has catalogued 132 galaxies
brighter than $m_{v}\simeq 16$.  Radial velocities are available for
113 of the galaxies in this flux-limited catalog (i.e., a completeness
of $86\%$).  Among these 113 galaxies, we find 99 cluster members and
14 outliers, which implies that we should expect to find about 2 more outliers
among the 19 galaxies for which a radial velocity is unavailable.
Hence the sample of $99+19=118$ cluster galaxies can be considered a
magnitude-limited sample almost free of field contamination. We shall
distinguish between the latter sample of 118 A2634 members within the
central half degree region (hereinafter referred to as the ``HD sample'')
and the subset of that of 99 galaxies with known redshift
(hereinafter referred to as the ``HDR sample'').

Figure~10 contains the stripe density plot and velocity histogram
for the 99 galaxies in the HDR sample. For this sample
 $V_{hel}=9151^{+84}_{-91}\,$ \kms and $\sigma=800^{+61}_{-52}\,$ \kms.
The results of the $W$- and $A^2$-tests indicate a more significant departure
from the gaussian model than for the TD sample, giving probabilities of a
gaussian
parent population of only $2\%$ and $3\%$, respectively. The velocity
distribution
appears also relatively skewed, but the $B_1$-test cannot reject the
gaussian hypothesis ($p(B_1)=0.29$). Similarly, the apparent bimodality of the
velocity histogram is not supported by the presence of significant gaps in the
velocity distribution. Figures 11 and 12 show, respectively, the velocity and
spatial distributions of the subsets of early- and late-type galaxies.
Although the velocity distribution of the E+S0 population has underpopulated
tails
(TI=0.89) and contains one highly significant gap of
146 \kms centered around 8770 \kms, all the statistical tests give markedly
non-significant rejection levels for the gaussian hypothesis (see Table 5).
For this subsample $V_{hel}=9240^{+79}_{-89}\,$ \kms and
$\sigma=661^{+74}_{-52}\,$ \kms. In contrast, the velocity distribution of
the late-type subsample is clearly non-gaussian, ---with the exception of
the $B_1$-test, the other tests reject the gaussian hypothesis at better than
the
$3\%$ significance level---, and multimodal, with suggestive
evidence of three possible kinematical subunits
with similar number of members. The small number of galaxies, however, hampers
the detection of highly significant gaps: only one gap of 548 \kms centered
around 8400 \kms falls within this category, the corresponding cumulative
probability of finding such a weighted gap
somewhere in this distribution being $1\%$. Two of the modes are located in the
lower and upper tails of the global velocity distribution, implying
that only $\sim 1/3$  of the late-type galaxies have velocities close
to the systemic velocity of the cluster. Accordingly, the main moments
of the velocity distribution of this late-type subset,
$V_{hel}=8897^{+206}_{-246}\,$ \kms and $\sigma=1030^{+116}_{-86}\,$ \kms
(see also Table 4), have values that are incompatible, within the adopted
uncertainties, with those of the early-type population.

Appart from the dynamical complexity of A2634, the HD sample also
reconfirms the presence of significant morphological segregation in this
cluster. Figures 11b and 12b show that the early-type galaxies dominate
in the densest regions where the late-type galaxies are almost absent.
Superposed to the spatial distribution of galaxies in both subsamples is the
adaptive kernel density contour map (Silverman 1986; Beers 1992) drawn using
the 118 galaxies in the HD sample. The adaptive kernel technique is a two-step
procedure which, after applying a pilot smoothing to estimate the local density
of
galaxies, uses a smoothing window the size of which decreases with increasing
local density, so the statistical noise in low-density regions can be supressed
without oversmoothing the high-density ones. Two noticeable subclumps can be
seen in the figures, the location of the secondary subclump being consistent
with
the northeast elongation seen in the X-ray image of Pi93. To check if the
apparent
substructures seen in the density map correspond to kinematically different
subunits, we have used different symbols for the galaxies according to three
velocity ranges:  7000--8700 (crosses), 8700--9500 (circles) and
9500--$11{,}500$ \kms (asterisks).  In both the early- and late-type subsamples
the velocities of the galaxies are totally independent of their sky positions.
The adaptive kernel map obtained using only the galaxies in the HDR sample
shows a close similarity in the form of the contours to the one presented here
for the HD sample. The agreement between the two maps implies that the detected
subclumps are not due to chance projections of field galaxies mistakenly
included
in the HD sample, and confirms a posteriori its fairness.

The above results suggest that the early- and late-type galaxy populations
of the HD sample have indeed very different spatial and kinematical
properties (a two-sided KS test gives a probability of only $1\%$ that
a difference as large, or larger, than the one observed between the two
velocity distributions can occur by chance).  The spatial and kinematical
distributions of velocities of early-type galaxies are compatible with them
being dynamically relaxed. On the other hand, the absence of spatial
segregation among the different kinematical subunits of the late-type
population and their narrow velocity dispersions, suggest that this
kinematical multimodality is more the signature of the infall of
individual galaxies or small
groups onto the cluster taking place mainly along the line-of-sight than that
of a recent merger with a dispersed subcluster moving in the plane of the sky.
It is also interesting to note that the correlation between velocity dispersion
and X-ray temperature in clusters of galaxies recently published by Lubin \&
Bahcall (1993), predicts a velocity dispersion for A2634 of $690\pm 140$ \kms,
very close to the value derived here for the early-type subsample.
Accordingly, we adopt the values of the velocity centroid and dispersion
of the early-type population in the HDR sample, i.e., $9240\pm 84$ \kms and
$661 \pm 63$ \kms, respectively (here the quoted uncertainties
correspond to the average of the $68\%$ bootstrap errors), as representative
of the whole cluster.  Note that these values are in excellent agreement with
those obtained by Pi93 for their restricted sample of 88 galaxies free from
substructure and spatial/kinematical outliers.

\subsection{The Velocity Offset of NGC 7720}  

The significance of the velocity offset of NGC 7720 with respect to the
systemic velocity of A2634 can be determined by the expression (Teague,
Carter, \& Grey 1990):
\begin{equation}  
S=\mid \Delta V_{off}\mid /(\epsilon_{clus}^{2}+\epsilon_{cD}^{2})^{1/2}\;,
\end{equation}
where $\Delta V_{off}$ is the (relativistically correct) velocity offset of the
cD,
and $\epsilon_{clus}$ and $\epsilon_{cD}$ are the corresponding uncertainties
in the velocity of, respectively, the cluster and the cD galaxy.
Using the velocity centroid adopted for A2634 and a heliocentric velocity of
NCG 7720 equal to $V_{\rm cD}=9154\pm 59$ \kms (Pi93), the corresponding
velocity offset is $\Delta V_{off}=-83\pm 99$ \kms (the errors in the
velocities of A2634 and NGC 7720 have been summed in quadrature). Thus,
we have $S=0.84$, implying the (quasi-)stationarity of the cD relative
to the cluster, in agreement with Pi93 result.

\subsection{Analysis of Substructure}  

Both the density contours obtained via the adaptive kernel method and the shape
of the velocity histograms (specially that of the late-type population) suggest
the existence of substructure in the inner regions of A2634. We ascertain the
statistical significance of such apparent substructure with the application of
specifically designed tests. In subsections 4.4.1 and 4.4.2 we scrutinize the
spatial
correlation properties of the galaxy distribution, while in subsection 4.4.3 we
concentrate on the detection of significant local deviations from the global
kinematics
of the cluster.

\subsubsection{Spatial Tests. The SSG Test}  

We first consider the test developed by Salvador-Sol\'e et al. (1993a;
hereinafter
referred to as the SSG test), which has been shown to be well suited for the
detection of small-scale substructure in systems with circular or elliptical
self-similar symmetry and with a small number of particles.

Let $s$ be the projected radial distance from the center of symmetry of
a (circularly symmetrized) cluster and $N(s)$ the projected number density
profile
of galaxies. The SSG test produces two different estimates of $N(s)$:
$N_{dec}(s)$ and $N_{dir}(s)$, which are, respectively, sensitive and
insensitive to the existence of correlation among galaxy positions
relative to the cluster background density; the difference among the
two is interpreted as an index of existence of substructure.

The estimate $N_{dec}(s)$ is obtained by inverting  (``deconvolution method'')
the relation:
\begin{equation} \label{dec}  
\Sigma(s)=\pi s(N_{dec}\ast N_{dec})(s)\;,
\end{equation}
where $\Sigma(s) \delta s$ is the number of pairs of galaxies with observed
separation between $s$ and $s + \delta s$, among the $N_{gal}(N_{gal}-1)/2$
pairs obtained from the $N_{gal}$ galaxies in the cluster sample.
$(N_{dec}\ast N_{dec})(s)$ is the autocorrelation of $N_{dec}(s)$, which,
for a radially symmetric function is also equal to its self-convolution.
The estimate $N_{dir}(s)$ is obtained by inverting (``direct method'') the
relation:
\begin{equation} \label{dir}  
\Pi(s)=2\pi sN_{dir}(s)\;,
\end{equation}
where $\Pi(s) \delta s$ is the number of galaxies at projected
distances between $s$ and $s + \delta s$ from the center of symmetry of
the galaxy distribution.  Contrary to $N_{dec}$, $N_{dir}$ does not
rely on the relative separations of galaxies, therefore it is
insensitive to the existence of correlation in galaxy positions.  In
practice, in order to use the full positional information in the data,
the inversion
of equations (\ref{dec}) and (\ref{dir}) relies on the cumulative forms
$\int_{s}^{\infty}\Sigma(x)\,dx$ and $\int_{s}^{\infty}\Pi(x)\,dx$,
rather than the distributions $\Sigma(s)$ and $\Pi(s)$ themselves, so that
\begin{equation}\label{n_dec} 
N_{dec}(s)={{\cal F}_1 \circ {\cal A}}\biggl[{{\cal A}\circ {\cal F}_{1}^{-1}}
\biggl( 2\int_{s}^{\infty}\Sigma(x)\,dx\biggr)\biggr]^{1/2}\;,
\end{equation}
and
\begin{equation}\label{n_dir}
N_{dir}(s)={{\cal F}_1 \circ {\cal A}}\biggl[{{\cal A}\circ {\cal F}_{1}^{-1}}
\biggl(\int_{s}^{\infty}\Pi(x)\,dx\biggr)\biggr]\;,
\end{equation}
where ${\cal F}_1$ and $\cal A$ stand, respectively, for the
one-dimensional Fourier and Abel transformations (Bracewell 1978), and
where the symbol ``$\circ$'' denotes the composition of functions. Before we
proceed further, a caveat is necessary. The profile $N_{dec}$ cannot
be inferred with an arbitrarily high spatial resolution. Because the
observables are not continuous functions, radial symmetry is always
broken at small
enough scales. This causes the argument of the square root in eq.~(\ref{n_dec})
to take negative values due to statistical fluctuations, which requires that we
filter out the highest spatial frequencies. This results in a final $N_{dec}$
profile convolved with a hamming window of smoothing size $\lambda_{min}$
corresponding to the minimum resolution-length that guarantees the fulfillment
of the radial symmetry condition. Although this additional smoothing is
unnecessary for $N_{dir}(s)$, it must be also applied in order not to introduce
any bias in the subsequent comparison of the two  profiles.

The significance of substructure is estimated from the null hypothesis that
$N_{dec}(s)$ arises from a poissonian realization of some unknown spatial
distribution of galaxies that led to {\it the observed distribution\/}
of radial distances.  The probability of this being the case is
calculated by means of the statistic
\begin{equation}\label{chi2}  
\chi^2 = {(N_{dec}(0)-N_{dir}(0))^2 \over{2S^2(0)}}\;,
\end{equation}
for one degree of freedom. In eq.~(\ref{chi2}), $N_{dec}(0)$ and
$N_{dir}(0)$ are the values of profiles $N_{dec}$ and $N_{dir}$ at
$s=0$. $S^2(s)$ is the radial run of the variance of
the $N_{dir}$ profiles of 100 simulated clusters convolved to the same
resolution-length $\lambda_{min}$ of the observed profile.
These simulated clusters are generated by the azimuthal scrambling of
the observed galaxy positions around the center of symmetry of the cluster,
i.e.,
by randomly shuffling between 0 and $2\pi$ the azimutal angle of each galaxy,
while
maintaining its clustercentric distance $s$ unchanged.

Figure 13a shows the projected density profiles $N_{dec}$ and $N_{dir}$ and
their associated standard deviations for the galaxies in the circularized HD
sample.
It is readily apparent from this figure that both profiles are equal within the
statistical uncertainties; this is confirmed by the inference from
eq.~(\ref{chi2}) that the probability that the two profiles are the
same is $60\%$. The resulting minimum resolution-length of
$0.27\rm\,Mpc$ puts an upper limit to the half-coherence length of any
possible clump that may remain undetected in the central regions of
A2634.  Notice that this value is much lower than the typical value of
$0.6\rm\,Mpc$ inferred by Salvador-Sol\'e, Goz\'alez-Casado, \& Solanes
(1993b) for the scale-length of the clumps detected in the Dressler \&
Shectman (1988a) clusters.

\subsubsection{Spatial Tests. The SSGS Test}  

Another useful estimate of the significance of subclustering is that
proposed by Salvador-Sol\'e et al. (1993b; hereinafter referred to as
the SSGS test), which can be considered a modification to the SSG test
and is based on the density in excess of neighbors from a random galaxy
in a cluster. The quantity
\begin{equation}\label{excess}  
 N_{gal}^{-1}(N_{dir}\ast N_{dir})(s)\bar \xi(s)\;,
\end{equation}
represents the probability in excess of random of finding one cluster
galaxy at an infinitesimal volume $\delta V$ located at a distance $s$
from a random cluster galaxy, per unit volume. In eq. (\ref{a2pcf}),
the ``average two-point correlation function'' statistic $\bar \xi(s)$,
a generalization for isotropic but inhomogeneus systems of the usual
two-point correlation function, is given by the expression:
\begin{equation}\label{a2pcf}  
\bar \xi(s)={(N_{dec}\ast N_{dec})(s)-(N_{dir}\ast N_{dir})(s)\over{(N_{dir}
\ast N_{dir})(s)}}\;,
\end{equation}
with
\begin{equation}\label{ndec2}  
(N_{dec}\ast N_{dec})(s)={{\cal F}_1\circ {\cal A}}\biggl[{{\cal A}\circ
{\cal F}_{1}^{-1}}\biggl( 2\int_{s}^{\infty}\Sigma(x)\,dx\biggr)\biggr]\;,
\end{equation}
and
\begin{equation}\label{ndir2}  
(N_{dir}\ast N_{dir})(s)={{\cal F}_1\circ {\cal A}}\biggl[{{\cal A}\circ
{\cal F}_{1}^{-1}}\biggl(\int_{s}^{\infty}\Pi(x)\,dx\biggr)\biggr]^{2}\;.
\end{equation}
In this case, the statistical significance of substructure is obtained
by checking the null hypothesis that $N_{dec}(s)$ arises from a
poissonian realization of some unknown spatial distribution of
galaxies, which is approximated by $N_{dir}(s)$. In practice,
the presence of substructure is estimated by the comparison of the
empirical function given by eq.~(\ref{excess}) with the mean and one
standard deviation of the same function obtained from a large number of
poissonian cluster simulations (i.e., both the radius and the azimuthal
angle of each galaxy are choosen at random) that reproduce the profile
$N_{dir}(s)$.
In the SSGS test, the use of poissonian simulations, instead of
angular scramblings, to estimate the statistical uncertainties translates
in a partial loss of sensitivity in the detection of substructure when
compared with the SSG test (see Salvador-Sol\'e et al. 1993b for further
details). This is compensated, however, by the fact that the two
functions $N_{dec}\ast N_{dec}$ and $N_{dir}\ast N_{dir}$ (eqs.
[\ref{ndec2}] and [\ref{ndir2}], respectively) can be inferred with any
arbitrarily
high spatial resolution, so there are no lower limits in the size of the
clumps that can be detected, as opposed to the SSG test.
Figure 13b shows the density in excess of neighbors (eq.~[\ref{excess}]) for
the galaxies in the circularized HD sample, calculated between its center and
twice
its maximum radius. The dashed line and the vertical solid lines
represent, respectively, the mean value of this function and its
$1\sigma$-error calculated from 200 poissonian simulations.  A
low-passband hamming filter leading to a resolution length of
$0.05\rm\,Mpc$ has been applied to attenuate the statistical
noise at galactic scales. As the ``signal'' is embedded in the ``noise'', it is
clear from
this figure that the SSGS  test does also not detect any significant
substructure
in the HD sample. Notice that both the SSG and the SSGS tests detect
substructure in the 50\% of the Dressler \& Shectman clusters
(Salvador-Sol\'e et al. 1993a,b).

\subsubsection{Kinematical Test}  

The Dressler \& Shectman (1988b) statistical test (hereinafter referred to
as the DS test) complements the above spatial correlation tests because
it is sensitive to kinematical substructure in the form of significant local
deviations from the global distribution of radial velocities. The DS test
is based on the comparison of the local velocity mean, $\bar V_{local}$,
and velocity dispersion, $\sigma_{local}$, associated with each galaxy
with measured radial velocity (calculated using that galaxy and its 10
nearest projected neighbors with measured velocities) with the mean
velocity, $\bar V$, and velocity dispersion, $\sigma$, of the entire sample.
For each galaxy, the deviation from the global values is defined by
\begin{equation} \label{delta}  
\delta^{2}={11\over{\sigma^{2}}}[(\bar V_{local}-\bar V)^{2}+(\sigma_{local}-
\sigma)^{2}]\;.
\end{equation}
The observed cumulative deviation $\Delta_{obs}$, defined as the sum of
the $\delta$'s for all the galaxies with measured radial velocities, is
the statistic used to quantify the presence of substructure. To avoid
the formulation of any hypothesis on the form of the velocity
distribution of the parent population, this statistic is calibrated by
Monte-Carlo simulations that randomly shuffle the velocities of the
galaxies while keeping fixed their observed positions. In this way any
existing correlation between velocities and positions is destroyed. The
significance of subclustering is then given in terms of the fraction of
simulated clusters for which their cumulative deviation $\Delta_{sim}$
is larger than $\Delta_{obs}$.

A visual judgment of the statistical significance of local deviations from the
global kinematics for the galaxies in our HDR sample can be done comparing
the plots in Figures 14a--d. Fig.~14a shows the spatial distribution of the
HDR sample galaxies (filled circles) superposed on the adaptive kernel density
contour map of the HD sample (dashed lines). The coordinates are measured with
respect to the center adopted for A2634. In Fig.~14b each galaxy is identified
with a circle whose radius is proportional to $e^\delta$ (with $\delta$
given above).  Hence, the larger the circle, the larger the deviation
from the global values (but beware of the insensitivity of  the
definition of $\delta$ given by eq.~[\ref{delta}] to the sign of the
deviations from the mean cluster velocity). The superposition of the
projected density contours (dashed lines) shows that, among the
subclumps seen in the adaptive kernel map, the small density
enhancement at plot coordinates (-15,3), and to less extent the density
enhancement at (8,12), are related with apparently large local
deviations from the global kinematics.  The remaining figures show two
of the 1000 Monte-Carlo models performed:  Fig.~14c corresponds to the
one whose $\Delta_{sim}$ is closest to the median of the $\Delta$'s of
all the simulations, while Fig.~14d corresponds to the simulation whose
$\Delta_{sim}$ is closest to the value of the upper quartile.
The comparison of Fig.~14b with these last two figures reveals that the
observed local deviations from the global kinematics in the HDR sample
are indeed statistically insignificant. This is corroborated by the
fact that the value of $\Delta_{sim}$ is larger than $\Delta_{obs}$ in
more than $71\%$ of the Monte-Carlo models.  Even if we run the same
test for the subset of late-type galaxies the value of $\Delta_{sim}$
is still larger than $\Delta_{obs}$ in the $57\%$ of the simulated
clusters.

Based on the results of the above spatial and kinematical analysis, we
consider the apparent clumpiness on the central regions of A2634 seen
in the kernel map as statistically insignificant (i.e., consistent with
poissonian fluctuations of the galaxy distribution). The kinematical
test has also provided quantitative confirmation of the fact that the
velocities of the late-type population are not segregated in the plane
of the sky, although we cannot exclude the possibility that an important
fraction of the spirals is located in loose groups superposed along the
line-of-sight. Therefore, if the merger
scenario is preferred in front of other interpretative frameworks, one
must conclude that the galaxy component of A2634 has suffered a faster
relaxation after the collision than the gaseous one.  Although
N-body/hydrodynamical simulations of cluster formation have revealed
the important role played by the shocks and turbulences generated in
mergers in the evolution of the ICM, there is no general agreement on
the timescales for the dynamical settlement of this component because
of its depence on the adopted initial conditions (e.g., Evrard 1990;
Roettiger, Burns, \& Loken 1993), making difficult to pronounce on how
likely to happen the above fact is. Perhaps, as suggested by the
simulations of Schindler \& M\"uller (1993), the two-dimensional
temperature distributions of the ICM, which are expected to be
observable with the next generation of X-ray satellites, will provide a
definitive answer to that question.

\section{Kinematics and Spatial Analysis of the Other Clusters and Groups}

In this section we investigate in some detail the spatial distribution and
kinematics of the galaxies associated with A2666, the two
clusters in the background of A2634 and the groups A2634--F and A2634--B.

\subsection{A2666} 
For A2666, we limit our analysis to the galaxies located within
$1\,r_A$ (1.16 degrees) of its center. In Figure 15, we show (a) the
velocity and (b) spatial distribution (b) of the 39 galaxies within
this region that have velocities in the range 6500--9500 \kms and are
not included in the TD sample of A2634.  The spatial distribution (plot
coordinates are given with respect to the adopted center for A2666; see
\S~3.2) shows the existence of a central compact subunit containing a
large fraction of the galaxies and a dispersed population significantly
separated from the central condensation. The biweight location and
scale of the whole sample are $V_{hel}=8118^{+81}_{-80}\,$ \kms and
$\sigma=533^{+126}_{-98}\,$ \kms. For A2666, $\Delta V_{LG}=239$ \kms
and $\Delta V_{CMB}=-363$ \kms. The velocity distribution appears to be
marginally consistent with the gaussian hypothesis for all the
statistical tests except the $A^2$-test, which rejects it at the $3\%$
level of significance. However, the most marked characteristic
exhibited by the velocity histogram of these 39 galaxies is the
presence of heavily populated tails (TI=1.56), strongly suggesting a
complex velocity field.  In an attempt to unravel whether the shape of
the velocity distribution is due to the presence of infalling galaxies
towards the central subunit, we have identified galaxies with
heliocentric velocities smaller than 7800 \kms with crosses, those in
the interval 7800--8600 \kms around the main velocity peak with circles
and those with velocities larger than 8600 \kms with asterisks. The
three kinematical subsets, however, appear well mixed in the sky. The
solid part of the histogram in Fig.~15a shows the distribution of
velocities of the 26 galaxies belonging to the central subunit (i.e.,
those within $0.5\,r_A$ of the cluster center).  The velocity
dispersion of the galaxies in this subsample is
$\sigma=380^{+121}_{-78}\,$ \kms, substantially smaller than that of
the whole sample (the systemic velocity remains practically unchanged;
see Table 4), but the tail index TI=1.75 is even larger. None of the
statistical tests can reject now the gaussianity of the parent
distribution (see Table 5). The heavily populated tails of the velocity
distribution of A2666 may also signal possible chance superpositions of
objects not physically bound to the cluster. The presence of
contaminating galaxies is to be expected both because of the
supercluster in which A2634 and A2666 are embedded and because of the
proximity of A2634 itself.  Notice, for instance, that all four objects
with radial velocity larger than 8600 \kms in Fig.~15a are well within
the $\Omega_0 = 0.5$ caustic of Fig.~5, and therefore might be
peripheral members of A2634.

\subsection{The Distant Clusters} 

In \S~3.1, the presence of two background clusters in the A2634
region was briefly discussed. Figure 16 shows the sky distribution of the
likely members of these two clusters within a $1\fdg5\!\times 1\fdg5$ region
around
the A2634 center.

A total of 40 galaxies within this region have radial velocities in the range
from $15{,}000$ to $21{,}000$ \kms, which is dominated by the rich background
cluster A2622. The biweight
location and scale of this subset of galaxies are, respectively,
$V_{hel}=18345^{+144}_{-150}\,$ \kms and
$\sigma=942^{+165}_{-109}\,$ \kms (see also Table 4), and the velocity
distribution appears to be fully consistent with a gaussian parent population
(Table 5). A2622 appears to be  embedded in a region of high galactic
density, perhaps a supercluster, that extends several core radii to the
southeast from its center.  The peak density is located at $\rm
R.A.\sim 23^h 32^m 40^s$, $\rm Dec.\sim 27\deg 4\arcmin$, approximately
$0\fdg9$ to the NW of A2634.  It coincides with one of the two
secondary peaks of diffused X-ray emission in the {\it ROSAT\/} PSPC
X-ray map of A2634 shown in Figure 17, and is very close to the
position of the galaxy that is probably the dominant galaxy of the
cluster (the radio galaxy 4C 27.53; see Riley 1975).

For the 31 galaxies with radial velocities in the range
$35{,}000$--$41{,}000$ \kms spread over the same $1\fdg5\!\times
1\fdg5$ region, we find $V_{hel}=37093^{+192}_{-156}\,$ \kms and
$\sigma=924^{+307}_{-265}\,$ \kms. In this case, the $W$-, $B_1$- and
$A^2$-tests reject the gaussian hypothesis at better than the $5\%$
level of significance, and the tail index, with a value of 1.51,
signals the presence of heavily populated tails in the velocity
distribution (similar results are obtained if the kinematical analysis
is restricted to the 16 galaxies in the central subunit). These results
suggest that this galaxy concentration  (hereinafter CL37) has a
complex velocity field, perhaps contaminated by outliers. Pi93
tentatively associated the secondary peak seen in their {\it
Einstein\/} IPC image of A2634 (see also Fig.~17) with a probable
background cluster at $\sim 37{,}000$ \kms, but could not further
elaborate on this idea for the lack of enough redshift measurements.
With twice as many redshifts, we can now confirm the presence of a rich
cluster of galaxies with its peak density located less than $0\fdg2$ to
the NW of A2634 and surrounded, as A2622, by several smaller galaxy
concentrations at the same distance. Based on CCD images that we have
obtained for the central region of A2634 and on the association of this
background cluster with a noticeable secondary peak of X-ray emission
seen in both the {\it Einstein\/} and {\it ROSAT\/} images, we identify
an elliptical galaxy at $\rm R.A. = 23^h 35^m 25\fs4$, $\rm Dec.=
26\deg 54\arcmin 36\arcsec$ and $cz_{hel}=37{,}322$ \kms as the most
probable central galaxy of this cluster.

The consistency of the association of these two background clusters
with secondary peaks in the X-ray image of A2634 can be estimated from
their predicted X-ray luminosities in the 2--$10\rm\,keV$ energy range,
calculated using their observed velocity dispersion and the
$L_X-\sigma$ relation (Edge \& Stewart 1991). For a cluster with a
velocity dispersion on the order of 900 \kms, $L_X \simeq 7\times
10^{44}\rm\,erg\,s^{-1}$, similar to the X-ray luminosity of a cluster
like Coma. This predicts total X-ray fluxes of $4.5\times
10^{-11}\rm\,erg\,cm^{-2}\,s^{-1}$ for A2622 and of $1.0\times
10^{-11}\rm\,erg\,cm^{-2}\,s^{-1}$ for CL37, which have to be compared with
the total flux of $1.2\times 10^{-11}\rm\,erg\,cm^{-2}\,s^{-1}$ measured
for A2634 (David et al. 1993). The negligible dependence on cluster
temperature of the fluxes measured in the {\it Einstein\/}
0.5--$3.0\rm\,keV$ and {\it ROSAT\/} 0.14--$2.24\rm\,keV$ energy bands
implies that the predicted ratios between the X-ray fluxes of the
background clusters and A2634 in the 2--$10\rm\,keV$ energy range
should be similar to those calculated in the energy bands of both the
{\it Einstein\/} IPC and {\it ROSAT\/} PSPC detectors.  The expectation
of a noticeable secondary peak associated with A2622 is, however, not
corroborated by Fig.~17. That is largely due to the fact that the image
was not flat-fielded, while the X-ray peak pressumably coincident with
A2622 is located close to both the edge of the {\it ROSAT\/} image and
the shadow of the mirror supports. Therefore, we accept the positional
coincidence as sufficient evidence for the identification of the two
secondary X-ray peaks in Fig.~17 at $13\arcmin$ and $50\arcmin$ to the
NW of the center of A2634 with the two background clusters A2622 and
CL37.

\subsection{The Groups A2634--F and A2634--B} 

In \S~3.2 we identified two groups in the neighborhood of A2634, labelled
respectively A2634--F and A2634--B. The former has 18 members (open squares
in Fig.~6) and a systemic velocity of $7546^{+65}_{-71}$ \kms; A2634--B has
17 members (open triangles in Fig.~6) and a systemic velocity of
$11{,}619^{+72}_{-90}$ \kms.  Both groups have very small velocity dispersions
(244 and 186 \kms, respectively), and are relatively spiral rich ($\sim 60\%$).
The two groups appear concentrated in small areas on the plane of  the
sky, dominating the galaxy counts on the east side of the cluster. The
galaxies associated with A2634--F are spread to the northeast of the
cluster center, at a median distance of $\sim 1\fdg1$, and  forming
perhaps two different subunits very close to each other, spatially and
kinematically. A2634--B is slightly more concentrated (although Pi93
also suggest that it may contain two different subunits) and located
slightly to the southeast of the cluster, at a median distance of
$0\fdg 6$ from its center. It is likely that both groups represent
separate dynamical entities in the vicinity of A2634.  In particular,
all the galaxies in A2634--B would have been automatically discarded
from cluster membership in the TD sample by the $3\,\sigma$-clipping
method of Yahil \& Vidal (1977). In \S~6.2, we shall study their
dynamical relation to the cluster.

\section{Dynamical Analysis}  

\subsection{Mass Estimates}  

The virial theorem is customarily used as the standard tool to estimate
the dynamical mass of galaxy clusters. Under the assumptions that the
cluster is a spherically symmetric system in hydrostatic equilibrium
and that the mass distribution follows closely that of the observed
galaxies independently of their luminosity, the total gravitating mass
of a cluster is given by
\begin{equation}  
M_{\rm VT} = {3\pi \over G} \sigma^2 R_H\;,
\end{equation}
where $\sigma$ is the line-of-sight velocity dispersion of the galaxies
taken here as the biweight scale estimate, and $R_H$ is the cluster
mean harmonic radius, defined as
\begin{equation}  \label{rharm}  
R_H = {D \over 2} N_{gal}(N_{gal}-1) \Bigr( \sum_i \sum_{j<i} {1 \over
\theta_{ij}}\Bigl)^{-1}\;,
\end{equation}
where $D$ is the cosmological cluster distance, $\theta_{ij}$ is the angular
separation
between galaxies $i$ and $j$, and $N_{gal}$ the total number of galaxies.

An alternative approach is to use the ``projected mass estimator''
(Bahcall \& Tremaine 1981; Heisler, Tremaine, \& Bahcall 1985)
\begin{equation} \label{pme}  
M_{\rm PM} = {32 \over{\pi GN_{gal}}} \sum_i  V_{i}^2 R_i\;,
\end{equation}
where $V_i$ is the observed radial component of the velocity of galaxy
$i$ with respect to the systemic velocity of the cluster, and $R_i$ is
its projected separation from the cluster center.  The numerical factor
in front of eq.~(\ref{pme}) assumes an isotropic distribution of galaxy
orbits. It is worth noting that the cluster masses obtained with these
two methods may underestimate the actual ones if the distribution of
matter is less concentrated than the light.

Mass estimates using the above two methods and their $68\%$
uncertainties (computed by means of the bootstrap technique and a
standard propagation of errors analysis) are listed in columns (9) and
(10) of Table 4 for A2634, A2666 and the two distant clusters.  We have
computed dynamical masses for the TD and HDR samples of A2634, and for
the galaxies within 1 and $0.5\,r_A$ from the center of A2666. In
addition, because of the clearly different spatial and kinematical
properties shown by the early- and late-type galaxies of A2634, we have
also derived mass estimates for the two populations.

For A2634, we can compare the different mass estimates for the HDR
sample given in Table~4 with the value of $2.5\times
10^{14}\rm\,M_{\odot}$ calculated for the central cluster region from
the observed  X-ray gas distribution (Eilek et al. 1984). The better
agreement is obtained with the mass estimates inferred for the
early-type population, while the late-type galaxies yield excessively
large values. This result adds further support to the idea, discussed
in \S~4.2, that the spiral galaxies of A2634 represent a young cluster
population, not yet in a relaxed dynamical state, and possibly with a
different distribution of orbits than the early-type galaxies.  A
similar difference between the mass estimates of the early- and
late-type galaxies is also present in the TD sample. Note that if the
spiral galaxies were all falling onto the cluster along purely radial
orbits, their average velocity would be on the order of $\sqrt{2}$
times the mean velocity of the relaxed cluster population. This higher
velocity would produce an overestimate of the cluster mass by a factor
of two, very close to the observed difference between the mass
estimates of the early- and late-type populations.

On the other hand, a rough estimate of the radius of the virialized
region for a typical rich cluster (Mayoz 1990) gives
$r_{vir}=2.48\rm\,Mpc$, which corresponds to $0\fdg 8$ at the distance
of A2634. This suggests that even some of  the early-type galaxies in
the TD sample might be part of an unrelaxed population. Consequently,
we adopt as the best estimate of the virial mass of A2634 that given by
the early-type population of the HDR sample.  Similarly, the spatial
distribution  of the galaxies associated with A2666 (Fig.~15b) suggests
that the best estimate of the virial mass of this cluster is given by
the galaxies within $0.5\,r_A$ of its center.

Table 4 also shows that the virial mass estimates for all the selected
samples are affected by smaller uncertainties and yield smaller values
than the projected mass estimator.  Indeed, Monte Carlo simulations of
clusters done by Heisler et al. (1985) show that this second method
tends to overestimate the dynamical masses if the samples are
contaminated by interlopers. This case probably applies for A2666,
where the projected mass estimates are one order of magnitude larger
than the virial masses, and for the two background clusters, because
the small size of the  corresponding samples makes them more
susceptible to contamination by outliers.  Notice, for instance, that
the ratio of masses between A2634 and A2666 ranges between  4 and 10
depending of the technique adopted. In the next section, we use only
the virial mass estimates in the study of the dynamical state of the
A2634\slash 2666 system.

\subsection{Two-Body Analysis}  

We now investigate by means of simple energy considerations whether the two
clusters A2634 and A2666 and the two small groups A2634--F and A2634--B in the
vicinity of A2634 form a gravitationally bound system.

In the framework of newtonian mechanics, a system of particles is
gravitationally bound if it has a negative total energy or,
equivalently, if  $v^2/2<GM_{tot}/r$, where $v$ represents the (system
averaged) global velocity dispersion, $r$ is the characteristic size of
the system (e.g., eq.~[\ref{rharm}]) and $M_{tot}$ is the total system
mass. In the particular case under study, as the masses of the groups
A2634--F and A2634--B are negligible with respect to the masses of the
two main clusters, the energetic analysis can be reduced to that of
three independent two-body systems: the system A2634\slash 2666, and
the two cluster/group systems A2634/2634--F and A2634/2634--B.  For a
system of two point masses the criterion for gravitational binding
expressed in terms of observable quantities is written as (Beers,
Geller, \& Huchra 1982):
\begin{equation}\label{ener}  
V_{rel}^2 R_{p}\leq 2GM_{tot}\sin^{2}\alpha \cos \alpha\;,
\end{equation}
where $V_{rel}=v\sin\alpha$ is the relative velocity between the two
components along the line-of-sight, $R_{p}= r\cos\alpha$ is their
projected separation, and  $\alpha$ is the
angle between the plane of the sky and the line joining the centers of
the two components.
Notice that eq.~(\ref{ener}) defines the region of bound orbits in the
$(\alpha,V_{rel})$-plane independently of the adopted value of $H_0$.

For the system A2634\slash 2666, the (relativistically correct)
relative velocity between the two clusters is $V_{rel}=1078\pm105$ \kms
(the quoted uncertainty is the rms of the average 68\% boostrap errors
associated with each cluster), while $R_p=9.1\rm\,Mpc$, computed from
the angular separation of their respective centers at the average
$z_{CMB}$ of these clusters. Figure 18a shows the variation of
$V_{rel}$ as a function of $\alpha$. Two different curves are drawn for
two different values of the total mass of the clusters. The solid curve
is drawn using our previous best estimates of the virial masses of the
two clusters, while for the dotted line we use the highest values
obtained in our calculations using the virial mass estimator (see
Table~4). The region of bound orbits is on the left of the curves. The
vertical lines correspond to the observed value of $V_{rel}$ (solid)
and its associated uncertainty (dashed). Fig.~18a suggests that these
two clusters are currently gravitationally unbound.

For the system A2634/2634--F, the relative velocity between the two
components is $V_{rel}= 1632\pm 108$ \kms, while for the system
A2634/2634--B, is $V_{rel}= 2292\pm 117$ \kms. From the estimated mean
angular separations between the groups and the cluster we have
$R_p=3.4\rm\,Mpc$ for  A2634/2634--F and $R_p=1.9\rm\,Mpc$ for
A2634/2634--B at the cosmic distance of A2634. Figures 18b and 18c show
the limits between the bound and unbound orbit regions in the
$(V_{rel}{,}\alpha)$-plane for these two systems. As in Fig.~18a, the
calculations done with the best virial mass estimate of A2634 are
represented by solid curves, while the dotted ones show the result of
using the highest value given by the virial mass estimator. In each
calculation the mass of the group is neglected with respect to the mass
of the cluster. These figures show that the mass of A2634 is too small
to form a bound system with the two groups A2634--F and A2634--B, a
result that validates, a posteriori, the membership criterion adopted
for A2634 in \S~3.2.

Therefore, from the results of the preceeding analysis, we  conclude
that it is unlikely that the whole system of clusters and groups around
A2634 is gravitationally bound.

\section{Conclusions}  

An extensive galaxy redshift survey around A2634 serves to reveal a
region of intricate topology at all scales. Besides the large-scale
structure associated with the PPS, we are able to identify in the
$6\deg$ wide field around A2634 several clusters and groups: (a) A2634;
(b) A2666; (c) two rich and massive background clusters, also detected
in the X-ray domain, located at respectively twice and four times the
redshift of A2634; and (d) A2634--F and A2634--B, two spiral rich
groups near A2634 at $\sim 7500$ and $\sim 11{,}500$ \kms,
respectively.  Simple energetic considerations suggest that the system
formed by A2634, A2666, A2634--F and A2634--B is gravitationally
unbound, despite the proximity among its members.

We also show that the conflicting results on the motion of A2634 with
respect to the CMB reported by Lucey et al. (1991a) may be in part {\it
but not fully\/} ascribed to the complex structure of the region and
lenient assignment of cluster membership to the galaxies in the adopted
samples. Other explanations, ---in addition to stricter TF and
$D_n-\sigma$ sample selection criteria---, are needed to solve the
problem.

The dynamical complexity of this region is also reflected in the
structure of the best sampled of its galaxy concentrations: A2634.
While the spatial, kinematical and dynamical properties of the
early-type population agree fairly well with those expected for a
relaxed system, the spiral galaxies are not only less concentrated to
the cluster core, but they also display strong evidence for
multimodality in their velocity distribution and dominate the
high-velocity tails, suggesting their recent arrival to the cluster. In
addition, spirals are virtually absent from the central parts of A2634,
a result similar to that obtained by Beers et al. (1992) for A400.
These results have two important implications.  First, the velocity
centroid of the (presumably virialized) innermost cluster region can be
miscalculated and/or its velocity dispersion overestimated if
(assymmetric) secondary infall plays an important role. It is therefore
advisable to use only the early-type galaxy population to derive these
two properties. Second, unless secondary infall is a recent event in
the life of these clusters, late-type newcomers must be efficiently
converted into early-type systems in order to explain the scarcity of
the former types in the innermost cluster regions. Based on these
considerations, we advocate the following choice of parameters for,
respectively,  A2634 and A2666: systemic velocities of 9240 and 8134
\kms, velocity dispersions of 661 and 380 \kms, and virial masses of
$5.2\times 10^{14}$ and $0.4\times 10^{14}\rm\,M_{\odot}$.

The clumpiness shown by the galaxy number density map of the central
regions of A2634 and the multimodal velocity distribution of the
late-type population, are investigated as possible signatures of a
recent collision of this cluster with a large subunit. Statistical
tests for substructure find, however, no significant evidence of
clumpiness in the galaxy component of A2634 that can corroborate the
ocurrence of a merger in the plane of the sky, but we cannot exclude
the existence of smaller, loose groups of spirals, unlikely associated
with dense ICMs, spatially superposed. This result indicates that the
structure and kinematics of A2634 reflect the continuous infall of
individual galaxies or small groups onto the cluster along the
line-of-sight, rather than the recent merger of two comparable subunits
moving in the plane of the sky.

\acknowledgments

The authors would like to thank Timothy Beers for kindly providing the
program ROSTAT used in the kinematical analysis and the software for
the calculation of the adaptive kernel density maps. We are also
indebted to Eduardo Salvador-Sol\'e and Guillermo Gonz\'alez-Casado who
developed the basic source code used in the two spatial tests of
substructure, and to Daniel Golombek for assistance in extracting
images from the ``Palomar Quick Survey'' at the STScI.  Ginevra
Trinchieri, Bill Forman and Alex Zepka provided valuable advice about
the {\it ROSAT\/} image of A2634, which was captured from the {\it
ROSAT\/} master data base maintained in the public domain by NASA.  MS
greatly benefited from many enlightening and entertaining discussions
with Enzo Branchini.  JMS acknowledges support by the United
States-Spanish Joint Committee for Cultural and Educational Cooperation
and the Direcci\'on General de Investigaci\'on Cient\'{\i}fica y
T\'ecnica through Postdoctoral Research Fellowships.  This work was
supported by grants AST--9115459 to RG, and AST--9023450 and
AST--9218038 to MPH.

\newpage

\newpage
\begin{figure}  
\caption{Large-scale spatial distribution for all the galaxies with measured
radial
velocities in a region of the Pisces-Perseus Supercluster centered on
A2634.}
\end{figure}
\begin{figure}  
\caption{Wedge diagram (R.A. vs $cz_{hel}$) of all the galaxies included in
Fig.~1 with
velocities smaller than $16{,}000$ \kms. A2634 is the elongated galaxy
concentration seen
in the middle of the diagram.}
\end{figure}
\begin{figure}  
\caption{Spatial distribution of galaxies with known redshift in the
$6\deg\!\times 6\deg$ field around A2634's center. Galaxies with
$m_{pg}\le 15.7$ are identified with filled circles, while open circles
are used for those fainter than that limit.}
\end{figure}
\begin{figure}  
\caption{Radial velocity histogram up to $45{,}000$ \kms for the galaxies in
Fig.~3. Bins are 500 \kms in width.}
\end{figure}
\begin{figure}  
\caption{Heliocentric velocity vs. angular separation for all galaxies
within $6\deg$ from the center of A2634. Unfilled symbols identify
probable A2666 members in the range $6500\le V_{hel}\le 9500$ \kms. Solid lines
are
the predicted caustics for
three different values of $\Omega_{0}$.}
\end{figure}
\begin{figure}  
\caption{(a) Heliocentric velocity vs. angular separation for the
galaxies in the inner $2\deg$ around the center of A2634. Solid lines are
the predicted caustics for
three different values of $\Omega_{0}$. (b)
Corresponding spatial distribution. In both figures large open squares identify
the 18
galaxies associated with A2634--F, while large open triangles identify the
17 members of A2634--B. Large asterisks and crosses mark,
respectively, foreground and background galaxies outside the caustic
$\Omega_{0}=0.5$ and not belonging to any of these two groups. Filled circles
identify
bona fide members of A2634. }
\end{figure}
\begin{figure}  
\caption{Stripe density plot and velocity histogram for all the galaxies in the
A2634 TD sample. Arrows mark the location of the most significant weighted
gaps in the velocity distribution (see text).}
\end{figure}
\begin{figure}  
\caption{(a) Stripe density plot and velocity histogram of the
early-type galaxies in the TD sample. Arrows mark the location of
the most significant weigthed gaps in the velocity distribution. (b) Sky
distribution of the galaxies in this subsample. Crosses
identify galaxies with $cz_{hel}<8700$ \kms, while open circles those with
$cz_{hel}\ge 8700$ \kms. Spatial coordinates are relative to the adopted center
for
A2634.}
\end{figure}
\begin{figure}  
\caption{Same as Fig.~8 but for the late-type galaxies in the TD
sample.}
\end{figure}
\begin{figure}  
\caption{Same as in Fig.~7 but for the all galaxies in the A2634's HDR sample.}
\end{figure}
\begin{figure}  
\caption{(a) Stripe density plot and velocity histogram of the
early-type galaxies in the HDR sample.  Arrows mark the location of the
most significant weigthed gaps in the velocity distribution. (b) Spatial
distribution of the galaxies in this subsample. Crosses
identify galaxies with $cz_{hel}<8700$ \kms, open circles galaxies with
$8750\le cz_{hel}\le 9500$ \kms and asterisks those with $cz_{hel}>9500$ \kms.
Solid lines are equally spaced contours of the adaptive kernel map drawn from
the galaxies in the HD sample. The contours range from $5.22\times
10^{-4}$ to $1.08\times 10^{-2} \rm\,galaxies\,arcmin^2$. The initial
smoothing scale is set to $0.5\rm\,Mpc$.
Spatial coordinates are relative to the center adopted for A2634 (see text).}
\end{figure}
\begin{figure}  
\caption{Same as in Fig.~11 but for the HDR late-type galaxies subsample.}
\end{figure}
\begin{figure}  
\caption{Results of the spatial tests for substructure. (a) The SSG test:
projected number density profiles
$N_{dec}$ (dashed line) and $N_{dir}$ (solid line) of
the galaxies in the HD sample; vertical bars give an estimate of the
$1\sigma$-error of these profiles (see text). (b) The SSGS test: density in
excess of neighbors from a random cluster galaxy for the HD sample
compared with the mean value (dashed line) and $1\sigma$-error
(vertical solid lines) of this function.}
\end{figure}
\begin{figure}  
\caption{Local deviations from the global kinematics for the the
galaxies galaxies in the HDR sample as measured by the DS test. (a)
Spatial distribution of the HDR sample galaxies (filled circles)
superposed on the adaptive kernel density contour map of the HD sample
(dashed lines).
(b) The position of galaxies are marked with open circles which radius
scales with their local deviation $\delta$ from the global kinematics
(see text), from which the test statistic $\Delta_{obs}=\sum\delta's$
is calculated. The adaptive kernel contour map of the HD sample is
superposed (dashed lines). (c)--(d) Monte-Carlo models of the HDR
sample obtained after 1000 random shufflings of the velocities of the
galaxies. Panel (c) shows the model with a cumulative
deviation $\Delta_{sim}$ closest to the median of the
simulations. Panel (d) corresponds to the model whose
$\Delta_{sim}$ is closest to the value of the upper quartile. Coordinates are
measured with respect to the center adopted for
A2634.}
\end{figure}
\begin{figure}  
\caption{(a) Velocity histograms of the
galaxies associated with A2666 located within $1\,r_A$ of its center (open) and
within $0.5\,r_A$ (solid). (b) Corresponding spatial distribution. Crosses
identify galaxies with $cz_{hel}<7800$ \kms, open circles galaxies with
$7800\le cz_{hel}\le 8600$ \kms and asterisks those with $cz_{hel}>8600$ \kms.
Spatial coordinates are relative to the center adopted for A2666 (see text).}
\end{figure}
\begin{figure}  
\caption{Spatial distribution in a $1\fdg5\!\times 1\fdg5$ region around
the A2634 center of the galaxies with heliocentric radial velocities between
$15{,}000$ and $21{,}000$ \kms (crosses), and between $35{,}000$ and
$41{,}000$ \kms (open circles). The large central cross marks the position of
the center of A2634.}
\end{figure}
\begin{figure}  
\caption{Raw, uncalibrated {\it ROSAT\/} PSPC X-ray map of A2634. The image has
not been flat-fielded to
correct for any gain variation across the PSPC field of view, and for the
shadowing of the mirror
support structure. The lowest contour level corresponds to 0.7 counts per
$15\arcsec\times 15\arcsec$
pixel; the remaining contours correspond to successive increments of $10\%$ on
the intensity. The diffuse X-ray emission associated with the two distant
clusters CL37 and A2622 can be seen superposed to that of A2634 at,
respectively, $13\arcmin$ and $50\arcmin$ to the NW of the cluster center.}
\end{figure}
\begin{figure}  
\caption{(a)--(c) The bound- and unbound-orbit
regions in the $(V_{rel}{,}\alpha)$-plane. Two
different curves are drawn for two different values of the total mass of
the corresponding systems. The solid line is drawn using the best virial mass
estimates, while the dotted line is drawn using the highest values given by the
virial mass estimator for the two-body systems under consideration (see text).
The region of bound orbits is on the left of the
curves. The vertical lines represent the observed value of $V_{rel}$
(solid) and its associated $68\%$ uncertainty (dashed).}
\end{figure}

\begin{references}
\reference Aaronson, M., Bothun, G.D., Mould, J.R., Huchra, J.P., Schommer,
R.A., \&
Cornell, M.E. 1986, \apj, 302, 536
\reference Abell, G.O. 1958, \apjs, 3, 211
\reference Bahcall, J.N., \& Tremaine, S. 1981, \apj, 244, 805
\reference Batuski, D., \& Burns, J.O. 1985, \apj, 299, 5
\reference Bautz, L.P., \& Morgan, W.W. 1970, \apjl, 162, L149
\reference Beers, T.C. 1992, in Statistical Challenges in Modern Astronomy, ed.
E.D.
Feigelson, \& G.J. Babu (New York: Springer-Verlag), 111
\reference Beers, T.C., Flynn, K., \& Gebhardt, K 1990, \aj, 100, 32
\reference Beers, T. C., Gebhardt, K., Huchra, J.P., Forman, W., Jones, C., \&
Bothun,
G.D. 1992, \apj, 400, 410
\reference Beers, T.C., Geller, M.J., \& Huchra, J.P. 1982, \apj, 257, 23
\reference Bertola, F., \& Perola, G.C. 1973, Astrophys. Lett., 14, 7
\reference Bird, C.M., \& Beers, T.C. 1993, \aj, 105, 1596
\reference Bracewell, R. 1978, The Fourier Transform and its Applications (New
York:
McGraw-Hill)
\reference D'Agostino, R.B. 1986, in Goodness of Fit Techniques, ed. R.B.
D'Agostino \& M.A. Stephens (New York: Dekker), 367
\reference Danese, L., De Zotti, G., \& di Tullio, G. 1980, \aap, 82, 322
\reference David, L.P., Slyz, A., Jones, C., Forman, W., Vrtilek, S.D., \&
Arnaud, K.A. 1993, \apj, 412, 479
\reference Davies, R.L., Burstein, D., Dressler, A., Faber, S.M., Lynden-Bell,
D., Terlevich, R.J., \& Wegner, G. 1987, \apjs, 64, 581
\reference de Lapparent, V., Geller, M.J., \& Huchra, J.P. 1989, \apj, 343, 1
\reference Dressler, A. 1980a, \apj, 236, 351
\reference Dressler, A. 1980b, \apjs, 42, 565
\reference Dressler, A., Lynden-Bell, D., Burstein, D., Davies, R.L., Faber,
S.M.,
Terlevich, R.J., \& Wegner, G. 1987, \apj, 313, 42
\reference Dressler, A., \& Shectman, S.A. 1988a, \aj, 95, 284
\reference Dressler, A., \& Shectman, S.A. 1988b, \aj, 95, 985
\reference Edge, A.C., \& Stewart, G.C. 1991, \mnras, 252, 428
\reference Eilek, J.A., Burns, J.O., O'Dea, C.P., \& Owen, F.N. 1984,
\apj, 278, 37
\reference Evrard, A. 1990, \apj, 363, 349
\reference Giovanelli, R., \& Haynes, M.P. 1985, \aj, 90, 2445
\reference Giovanelli, R., \& Haynes, M.P. 1989, \aj, 97, 633
\reference Giovanelli, R., \& Haynes, M.P. 1993, \aj, 105, 1271
\reference Giovanelli, R., Haynes, M.P., \& Chincarini, G. 1986a, \apj, 300, 77
\reference Giovanelli, R.,  Haynes, M.P., Myers, S.T., \& Roth, J. 1986b, \aj,
92, 250
\reference Giovanelli, R., Scodeggio, M., Solanes, J.M., Haynes, M.P.,
Arce, H., \& Sakai, S. 1994, in preparation
\reference Hamilton, D., Oke, J B., Carr, M.A., Cromer, J., Harris, F.H.,
Cohen, J., Emery, E., \& Blake\'e, L. 1993, \pasp, 105, 1308
\reference Hagfors, T., et al. 1989, proposal submitted to the NSF
\reference Heisler, J., Tremaine, S., \& Bahcall, J.N. 1985, \apj, 298, 8
\reference Jones, C., \& Forman, W. 1984, \apj, 276, 38
\reference Kent, S.M., \& Gunn, J.E. 1982, \aj, 87, 945
\reference Lubin, L.M., \& Bahcall, N.A. 1993, \apj, 415, L20
\reference Lucey, J.R., Bower, R.G., \& Ellis, R.S. 1991b, \mnras, 249, 755
\reference Lucey, J.R., Gray, P.M., Carter, D., \& Terlevich, R.J. 1991a,
\mnras, 248, 804
\reference Matthews, T.A., Morgan, W.W., \& Schmidt, M. 1964, \apj, 140, 35
\reference Mayoz, E. 1990, \apj, 359, 257
\reference Nilson, P. 1973, Uppsala General Catalogue of Galaxies (Uppsala
Astron. Obs. Ann., Vol. 6)
\reference Oke, J.B., \& Gunn, J. E. 1982, \pasp, 94, 586
\reference Pinkney, J., Rhee, G., Burns, J.O., Hill, J.M., Oegerle, W.R.,
Batuski, D., \& Hintzen, P. 1993, \apj, 416, 36 (Pi93)
\reference Reg\"os, E., \& Geller, M J. 1989, \aj, 98, 755
\reference Riley, J.M. 1975, \mnras, 170, 53
\reference Riley, J.M., \& Branson, N.J.B.A. 1973, \mnras, 164, 271
\reference Rosenberger, J.L., \& Gasko, M. 1983,  in Understanding Robust and
Exploratory Data analysis, ed. D.C. Hoaglin, F. Mosteller, \& J.W. Tukey (New
York: Wiley), 297
\reference Roettiger, K., Burns, J., \& Loken, C. 1993, \apjl, 407, L53
\reference Salvador-Sol\'e, E., Sanrom\'a, M., \& Goz\'alez-Casado, G. 1993a,
\apj,
402, 398
\reference Salvador-Sol\'e, E., Goz\'alez-Casado, G., \& Solanes, J.M. 1993b,
\apj, 410, 1
\reference Salvador-Sol\'e, E., \& Solanes, J.M. 1993, \apj, 417, 427
\reference Sastry, G., \& Rood, H. 1971, \apjs, 72, 75
\reference Schechter, P. 1976, \apj, 203, 297
\reference Schindler, S.,\& M\"uller, E. 1993, \aap, 272, 137
\reference Scott, J.S., Robertson, J.W., \& Tarenghi, M. 1977, \aap, 59, 23
\reference Silverman, B.W. 1986, Density Estimation for Statistics and Data
Analysis
(London: Chapman and Hall)
\reference Smoot, G.F., et al. 1991, \apjl, 371, L1
\reference Sodr\'e, L., Capelato, H.V., \& Steiner, J.E. 1989, \aj, 97, 1279
\reference Teague, P.F., Carter, D., \& Grey, P.M. 1990, \apjs, 72, 715
\reference Tonry, J., \& Davis, M. 1979, \aj, 84, 1511
\reference Tully, R.B., \& Fisher, J.R. 1977, \aap, 54, 661
\reference Tully, R.B., \& Shaya, E.J. 1984, \apj, 281, 31
\reference van Haarlem, M.P., Cay\'on, L., Guti\'errez de la Cruz, C.,
Mart\'{\i}nez-Gonz\'alez, E., Rebolo, R. 1993, \mnras, 264, 71
\reference Wegner, G., Haynes, M.P., \& Giovanelli, R. 1993, \aj, 105, 1251
\reference Whitmore, B.C., \& Gilmore, D.M. 1991, \apj, 367, 64
\reference Yahil, A., \& Vidal, N.V. 1977, \apj, 214, 347
\reference Zabludoff, A.I., Huchra, J.P., \& Geller, M.J. 1990, \apjs, 74, 1
\reference Zwicky, F., Herzog, E., Wild, P., Karpowicz, M., \& Kowal, C.
1961--1968, Catalogue of Galaxies and Clusters of Galaxies (Pasadena:
California Institute of Technology)
\end{references}
\end{document}